\documentclass[letter]{aa}

    \usepackage{graphicx} \usepackage{txfonts} \usepackage{amsmath} \usepackage{booktabs} \usepackage{array} \usepackage{multicol} \usepackage{natbib} \usepackage[colorlinks=true,allcolors=blue]{hyperref}

    \makeatletter \newenvironment{appendixfigure} {\par\begingroup\noindent\begin{minipage}{\textwidth}\def\@captype{figure}\centering} {\end{minipage}\par\endgroup} \newenvironment{appendixtable} {\par\addvspace{\intextsep}\begingroup\noindent\begin{minipage}{\linewidth}\def\@captype{table}\centering} {\end{minipage}\par\endgroup\addvspace{\intextsep}} \makeatother

\begin{document}

    \title{The Neptunian ridge through the lens of TESS} \titlerunning{The Neptunian ridge through the lens of TESS}

    \author{A.~Castro-Gonz\'alez\thanks{Corresponding author: \protect\texttt{amadeo.castro-gonzalez@unige.ch}}} \authorrunning{A.~Castro-Gonz\'alez} \institute{Observatoire Astronomique de l'Universit\'e de Gen\`eve, Chemin Pegasi 51b, 1290 Versoix, Switzerland} \date{Received 3 November 2026 / Accepted 1 December 2026}

    \abstract {The Neptunian desert and savanna are separated by a ridge, an overdensity of planets at orbital periods of 3--6~d identified with \textit{Kepler}. The ridge, together with the analogous hot Jupiter pileup at larger radii, has been linked to the accumulation of successful outcomes of high-eccentricity migration (HEM).} {We aim to characterise the ridge independently with TESS and compare its contrast with estimates from earlier \textit{Kepler} studies.} {We analyse TESS occurrence posteriors for FGK main-sequence stars, quantify the ridge excess over both neighbouring regions, and compare six occurrence profiles through leave-one-out cross-validation and Bayesian evidence.} {For planets with radii 5--8~$\mathrm{R}_\oplus$, the ridge has twice the occurrence density of the denser neighbouring region, with 99.9\% posterior support for this excess (one-sided Gaussian equivalent of $3.1\sigma$), in agreement with the contrast estimates from the discovery study and earlier \textit{Kepler} occurrence studies. The model comparison strongly favours profiles peaking within the ridge over a single power law. Predictive performance and Bayesian evidence both favour the lognormal, suggesting a narrower inner wing and a more extended outer wing, although the evidence for asymmetry remains weak. In the same period window, the analysis also recovers the hot Jupiter pileup (10--16~$\mathrm{R}_\oplus$) with a contrast consistent with that of the ridge and excludes an analogous excess among sub-Neptunes \mbox{(2--4~$\mathrm{R}_\oplus$)} with $3.4\sigma$ posterior support.} {TESS provides an independent detection of the Neptunian ridge. The possible asymmetry suggested by the data would be compatible with HEM, in which tidal disruption could produce a sharper inner edge and variations in circularisation efficiency a more gradual outer decline. Establishing whether the ridge has this shape requires more precise occurrence measurements, with PLATO's expected planet yield offering a promising opportunity to help unveil its morphology.}


    \maketitle \nolinenumbers

    \section{Introduction} \label{sec:introduction}

    Exoplanet surveys have identified a pronounced deficit of planets at very short orbital periods, known as the Neptunian desert \citep[e.g.][]{SzaboKiss2011}. Its origin is still debated, with atmospheric erosion \citep{EhrenreichDesert2011,Bourrier2018,OwenLai2018} and tidal disruption during dynamical migration \citep{MatsakosKonigl2016,CastroGonzalez2026TOI426,CastroGonzalez2026} among the main proposed mechanisms. Using occurrence rates derived from \textit{Kepler}, \citet{CastroGonzalez2024} identified a sharp rise in Neptunian occurrence at the desert edge, followed by a decline towards longer periods. This produces an overdensity at \mbox{$\approx$~3--6~d}, termed the Neptunian ridge, which separates the short-period desert from a moderately populated region at longer periods, known as the Neptunian savanna \citep{Bourrier2023}. The ridge's overdensity resembles that of the hot Jupiter pileup \citep{Udry2003}, raising the possibility that related processes may shape these populations \citep{CastroGonzalez2024}.

    Both the pileup and the ridge appear as an increase followed by a near plateau when the occurrence is expressed per unit $\ln P$ \citep{Kunimoto2020,Zink2023,ZinkHoward2023,Cui2026Ridge}. However, these excesses in occurrence per day are recognised as distinct features because their planets also differ systematically in key physical properties and orbital architectures from those on wider orbits. Jovian planets in the pileup typically follow nearly circular orbits, whereas those just beyond it have a broader eccentricity range \citep{Correia2020}, consistent with the pileup reflecting the circularised outcomes of high-eccentricity tidal migration (HEM; \citealt{FabryckyTremaine2007}). Jupiters in the pileup also show a pronounced deficit of nearby low-mass companions relative to the Jupiters beyond it \citep{Huang2016,WuRiceWang2023}, an expected consequence of HEM \citep{Mustill2015}. A similar architectural dichotomy is seen in the Neptunian regime, where ridge planets are typically isolated whereas savanna planets often reside in compact multi-planet systems, with a marked transition near the ridge's outer edge \citep[e.g.][]{Almenara2026,Thomas2026}. In addition, the transition between the ridge and the savanna is accompanied by a substantial change in bulk density, with ridge planets reaching high values ($\approx$~1.5--2.5~$\mathrm{g\,cm^{-3}}$) that are largely absent among savanna Neptunes \citep{CastroGonzalez2024TOI5005}, a contrast that might reflect tidally driven envelope loss during HEM \citep[e.g.][]{HallattMillholland2026,CastroGonzalez2026}. Recently, \citet{Abhilash2026} interpreted the larger masses and inferred envelope mass fractions of ridge planets as evidence for gas-rich formation at wider separations followed by HEM. Overall, these marked transitions point to distinct evolutionary histories within and beyond the ridge and pileup, with HEM as a leading explanation for both overdensities \citep{Wu2007,CastroGonzalez2026,Zanazzi2026}.

    In this Letter, we revisit the Neptunian ridge using the occurrence posteriors released by \citet{Cui2026} for Sectors~1--55 of the Transiting Exoplanet Survey Satellite (TESS; \citealt{Ricker2015}). We describe the data and analysis in Sect.~\ref{sec:data}, present the results in Sect.~\ref{sec:results}, discuss them in Sect.~\ref{sec:discussion}, and conclude in Sect.~\ref{sec:conclusions}.

    \begin{figure*}[t] \centering \includegraphics[width=\textwidth]{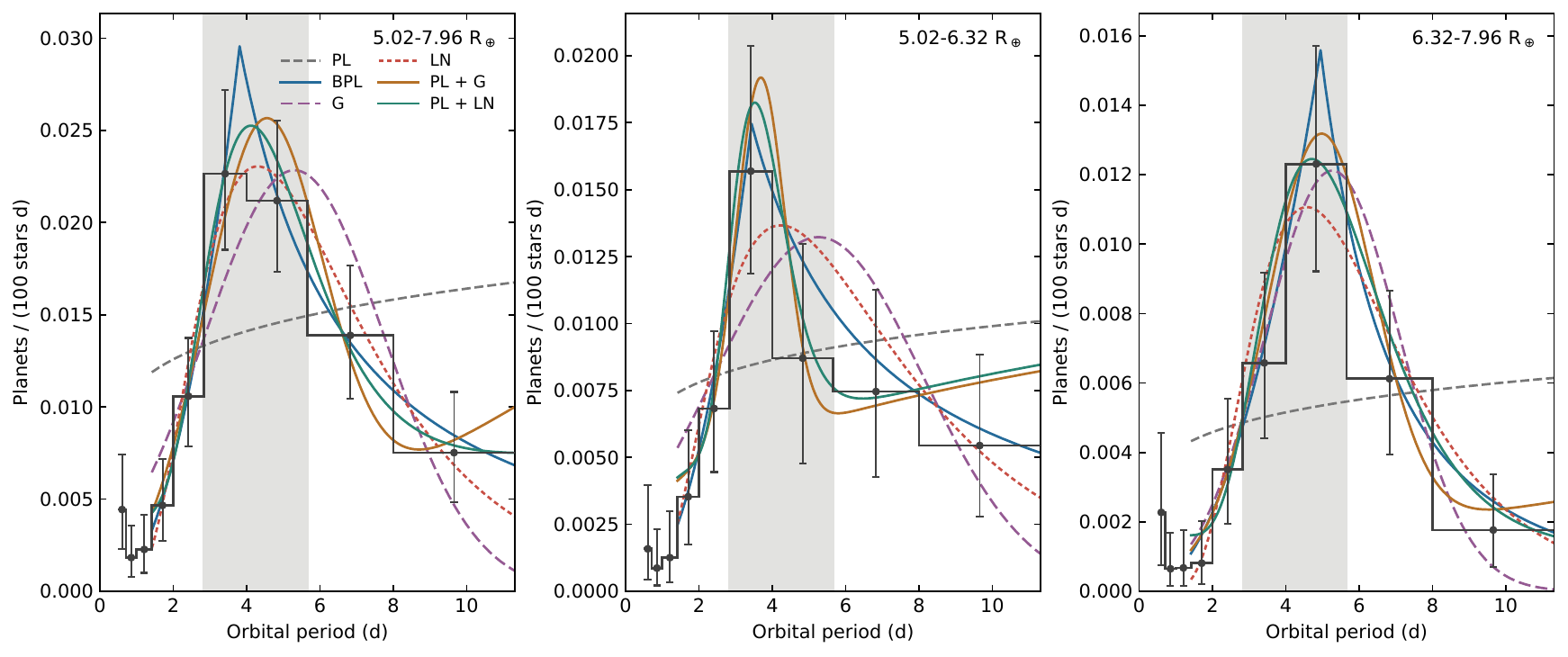} \caption{Occurrence density for 5.02--7.96~$\mathrm{R}_\oplus$ (left) and its two radius intervals, 5.02--6.32 (centre) and 6.32--7.96~$\mathrm{R}_\oplus$ (right). Histograms and filled circles show posterior medians, with error bars spanning the central 68\% credible intervals. Grey shading marks the adopted 2.83--5.66~$\mathrm{d}$ ridge interval (Sect.~\ref{sec:samples}). Curves show the highest-likelihood sampled PL (grey dashed), BPL (blue solid), G (purple long-dashed), LN (red short-dashed), PL+G (orange solid), and PL+LN (green solid) profiles (Sect.~\ref{sec:profilemethod}).} \label{fig:fgk} \end{figure*}

    \section{Data and analysis} \label{sec:data}

    \subsection{Occurrence data and period--radius selections} \label{sec:samples}

    We analyse the occurrence-rate posteriors released by \citet{Cui2026} for their combined sample of FGK main-sequence stars. The archive\footnote{\url{https://doi.org/10.5281/zenodo.17804280}.} contains 6960 posterior samples on a $10\times10$ grid spanning orbital periods of 0.5--16~$\mathrm{d}$ and planetary radii of 2--20~$\mathrm{R}_\oplus$, with each value representing the occurrence integrated over one period--radius cell. To explore possible differences across stellar types, we also analyse the posteriors for F, G, and K stars separately. These samples use the same period--radius grid as the combined FGK sample and are publicly available through the authors' occurrence portal.\footnote{\url{https://cuikaiming.com/TESS-SPOC-Occurrence-Rate/}. Accessed in October 2026.}

    The period bin edges are $P_j=0.5\times2^{j/2}$~$\mathrm{d}$, with $j=0,\ldots,10$. \citet{CastroGonzalez2024} identified the ridge at 3.2--5.7~$\mathrm{d}$, between the desert at shorter periods and the savanna at longer periods. Since our aim is to investigate whether the ridge has a higher occurrence than its adjacent regions, we select three intervals that closely represent the three regions, each comprising two period bins: $D=[1.41,2.83]$~$\mathrm{d}$ for the desert, $R=[2.83,5.66]$~$\mathrm{d}$ for the ridge, and $S=[5.66,11.31]$~$\mathrm{d}$ for the savanna. The central interval extends only slightly inward of the original ridge's inner boundary and nearly coincides with its outer boundary, providing a close match to the ridge's period range.

    The planetary radius bin edges are $R_j=2\times10^{j/10}\,\mathrm{R}_\oplus$, with $j=0,\ldots,10$. \citet{CastroGonzalez2024} detected the Neptunian ridge among planets with radii of 5.5--8.5~$\mathrm{R}_\oplus$. We therefore study the intervals 5.02--6.32 and 6.32--7.96~$\mathrm{R}_\oplus$, both separately and together. Although this selection extends slightly below the original lower radius boundary and omits its uppermost portion, the substantial overlap provides good coverage of the ridge's radius range. For completeness, we explore the hot Jupiter pileup (10.02--15.89~$\mathrm{R}_\oplus$), as well as the period distributions of sub-Neptunes (2--3.99~$\mathrm{R}_\oplus$) and planets in the transition regions at 3.99--5.02 and 7.96--10.02~$\mathrm{R}_\oplus$ (Appendix~\ref{app:populations}).

    \subsection{Occurrence densities and contrasts} \label{sec:method}

    The period intervals of the desert, ridge, and savanna have different widths, so a larger integrated occurrence can reflect a wider interval rather than a higher concentration of planets. To correct for this width effect, we divide the occurrence in each region by the length of its period interval: \begin{equation} h_s(B)=\frac{F_s(B)}{\Delta P_B}, \label{eq:occurrencedensity} \end{equation} where $B\in\{D,R,S\}$ denotes a region, $F_s(B)$ is its integrated occurrence in the posterior sample $s$, and $\Delta P_B$ is its period width.

    We first explore whether the ridge constitutes a local maximum in occurrence density relative to the desert and the savanna. To assess this, we compare its occurrence density with that of the denser of these two neighbouring regions: \begin{equation} J_s=\frac{h_s(R)}{\max[h_s(D),h_s(S)]}. \end{equation} A value $J_s>1$ means that the ridge exceeds both the desert and the savanna in occurrence density, establishing a local excess.

    To quantify the exterior contrast, we also calculate \begin{equation} C_s=\frac{h_s(R)}{h_s(S)}=\frac{2F_s(R)}{F_s(S)}. \end{equation} This ratio measures how much more densely populated the ridge is than the adjacent savanna. The two contrasts satisfy $J_s=C_s$ when the savanna is the denser neighbour and $J_s<C_s$ when the desert is denser, reflecting the requirement that a local excess must rise above both neighbouring regions.

    We calculate both contrasts for the 6960 posterior samples. We then obtain their posterior medians and central 68\% credible intervals, and use the fraction of samples above unity to estimate the posterior probability of an excess, $p_+$. To express the significance of the excess in sigma units, we convert this posterior probability to its one-sided Gaussian equivalent, $Z_{\rm eq}=\Phi^{-1}(p_+)$, where $\Phi$ is the standard normal cumulative distribution function.

    \subsection{Profile shape and model comparison} \label{sec:profilemethod}

    We also explore whether a power law, which cannot produce a local maximum, can describe the occurrence profile across the desert, ridge, and savanna. Since the period intervals scale successively by a factor of two, a power-law density $h(P)=AP^\beta$ predicts the same ratio of integrated occurrences from $D$ to $R$ and from $R$ to $S$. Therefore, $F(R)/F(D)=F(S)/F(R)$, giving $F(R)=\sqrt{F(D)F(S)}$ for any slope. To assess possible departures from a power law, we compare the ridge occurrence with this prediction in each posterior sample using the ratio \begin{equation} K_s=\frac{F_s(R)}{\sqrt{F_s(D)F_s(S)}}, \end{equation} where $K_s>1$ indicates that the ridge occurrence exceeds the power-law prediction.

    To further study the shape of the occurrence profile, we compare six models: a power law (PL), a broken power law (BPL), a Gaussian (G), a lognormal (LN), and a power law with a Gaussian (PL+G) or lognormal (PL+LN) component. The BPL can describe a maximum through a rising inner slope and a falling outer slope, while the Gaussian describes a symmetric peak and the lognormal an asymmetric peak with a longer outer wing. We compare their predictive performance using leave-one-out cross-validation (LOO) \citep{Vehtari2017} and assess their Bayesian support through differences in log evidence, $\Delta\ln\mathcal Z_{a,b}=\ln\mathcal Z_a-\ln\mathcal Z_b$, where positive values favour model $a$. We regard $1<\Delta\ln\mathcal Z<2.5$ as weak evidence, $2.5<\Delta\ln\mathcal Z<5$ as moderate evidence, and $\Delta\ln\mathcal Z>5$ as strong evidence. The statistical details of both analyses are given in Appendix~\ref{app:models}.

    \section{Results} \label{sec:results}

    For 5.02--7.96~$\mathrm{R}_\oplus$, the Neptunian ridge has twice the occurrence density of the denser neighbouring region, with $J=2.05^{+0.52}_{-0.44}$ (Fig.~\ref{fig:fgk}, left panel). The ridge exceeds both neighbouring regions in 6953 of 6960 posterior samples, giving $p_+=99.899\%$, corresponding to a Gaussian equivalent of $3.1\sigma$ ($Z_{\rm eq}=3.09$), supporting a local excess in occurrence per unit orbital period. Comparing the ridge with both neighbours gives only one fewer sample above unity than comparing it with the savanna alone, showing that the posterior support for a local excess is determined almost entirely by the ridge--savanna contrast. When analysed separately, the radius intervals 5.02--6.32 and 6.32--7.96~$\mathrm{R}_\oplus$ (Fig.~\ref{fig:fgk}, centre and right panels) both favour a local excess, with $J=1.62^{+0.71}_{-0.51}$ and $J=2.54^{+1.14}_{-0.77}$, respectively, although neither individually reaches $3\sigma$ posterior support. Table~\ref{tab:contrasts} lists the $J$ and $C$ contrasts, $p_+$, and $Z_{\rm eq}$ for all three radius selections. We also compare the ridge across stellar types, and the separate F-, G-, and K-star samples do not establish significant differences in contrast (Appendix~\ref{app:spectral}).

    \begin{table}[t] \setlength{\VSpaceAfterTabCaption}{3pt} \setlength{\VSpaceBeforeTabFoot}{0.5ex} \caption{Ridge contrasts with the denser neighbouring region ($J$) and the adjacent savanna ($C$).} \label{tab:contrasts} \centering\small \setlength{\tabcolsep}{4pt} \renewcommand{\arraystretch}{1.15} \begin{tabular*}{0.94\columnwidth}{@{\extracolsep{\fill}}clrrrr@{}} \hline\hline Contrast & Radius ($\mathrm{R}_\oplus$) & Ratio & $p_+$ (\%) & $Z_{\rm eq}$ & $N_{\leq1}$ \\ \midrule $J$ & 5.02--6.32 & $1.62^{+0.71}_{-0.51}$ & 89.885 & 1.28 & 704 \\ & 6.32--7.96 & $2.54^{+1.14}_{-0.77}$ & 99.483 & 2.56 & 36 \\ & 5.02--7.96 & $2.05^{+0.52}_{-0.44}$ & 99.899 & 3.09 & 7 \\ \addlinespace $C$ & 5.02--6.32 & $1.82^{+1.08}_{-0.63}$ & 91.983 & 1.40 & 558 \\ & 6.32--7.96 & $2.71^{+1.44}_{-0.87}$ & 99.626 & 2.68 & 26 \\ & 5.02--7.96 & $2.14^{+0.64}_{-0.49}$ & 99.914 & 3.13 & 6 \\ \bottomrule \end{tabular*} \tablefoot{Ratios are posterior medians with uncertainties spanning the central 68\% credible intervals. $p_+$ is the posterior fraction above unity, $Z_{\rm eq}$ its one-sided Gaussian equivalent, and $N_{\leq1}$ the number of samples at or below unity out of 6960.} \vspace{-5pt} \end{table}

    The ridge occurrence also exceeds the power-law prediction for 5.02--7.96~$\mathrm{R}_\oplus$, with $K=2.38^{+0.57}_{-0.45}$ and $K>1$ in all 6960 posterior samples. For the separate 5.02--6.32 and 6.32--7.96~$\mathrm{R}_\oplus$ radius intervals, we obtain $K=1.97^{+0.85}_{-0.58}$ and $K=3.36^{+1.57}_{-1.01}$, respectively, also indicating a departure from a power law. The LOO comparison reinforces this departure, as the BPL improves predictive performance over the PL by $12.07\pm4.28$ in ELPD. The BPL fit places the break at $p_b=3.73^{+0.73}_{-0.61}$~$\mathrm{d}$, which lies within the ridge, with a rising inner slope $\beta_1=2.30^{+1.18}_{-0.84}$ and a falling outer slope $\beta_2=-1.30^{+0.53}_{-0.74}$, describing a density maximum that complements the excess quantified by $J$ and $C$. Unequal slope magnitudes imply a local asymmetry around the maximum. Conditional on a maximum, the joint posterior gives a 78\% probability that the inner slope has the larger magnitude, hinting at a sharper inner rise than outer decline. To study this possible asymmetry further, we compare symmetric Gaussian and asymmetric lognormal peaks, both alone and with a PL background. Among all six models, the LN has the highest predictive score, improving on the symmetric G by $4.50\pm1.33$ in ELPD and more modestly on the BPL by $1.14\pm0.41$. The LN fit places its maximum at $\mu=4.35^{+0.37}_{-0.30}$~$\mathrm{d}$, also within the ridge. These results favour an asymmetric lognormal description of the ridge, with a longer outer wing. The LN also has a higher score than the G in both individual radius intervals. The fitted profiles are shown in Fig.~\ref{fig:fgk}, and Table~\ref{tab:profilecomparison} lists $\Delta\mathrm{ELPD}$ for all six models.

    We use Bayesian evidence to assess support for asymmetry while accounting for differences in model complexity. Both BPL and LN are strongly favoured over PL, with $\Delta\ln\mathcal Z_{\rm BPL,PL}=6.41$ and $\Delta\ln\mathcal Z_{\rm LN,PL}=8.15$. The LN has the highest evidence, with a weak advantage over the BPL ($\Delta\ln\mathcal Z_{\rm LN,BPL}=1.74$). With three parameters rather than four, it also offers a simpler description of the ridge. A direct comparison with the symmetric G, which also has three parameters, gives $\Delta\ln\mathcal Z_{\rm LN,G}=1.90$. Together, these results support the LN as a description of the ridge, with weak evidence for an asymmetric profile characterised by a narrower inner wing and a more extended outer wing. Table~\ref{tab:profilecomparison} compares $\Delta\ln\mathcal Z$ for all six models, with LN retaining the highest evidence and PL the lowest across the tested priors (Appendix~\ref{app:priors}).

    \section{Discussion} \label{sec:discussion}

    With a significant occurrence excess over both adjacent regions and a peaked profile favoured by model comparison, the Neptunian ridge is recovered in the TESS data, corroborating its original detection with \textit{Kepler} \citep{CastroGonzalez2024}. The TESS analysis also recovers the hot Jupiter pileup with a contrast consistent with that of the ridge, which suggests that processes such as HEM may produce similarly prominent concentrations in both populations over the same period range. In contrast, the same analysis excludes an analogous occurrence excess for sub-Neptunes, with $3.4\sigma$ posterior support. Appendix~\ref{app:logperiod} compares the ridge and pileup profiles with their transformed occurrence densities per unit $\ln P$, which both flatten as previously discussed for the pileup \citep[e.g.][]{Zink2023,ZinkHoward2023}.

    To explore the similarities and differences between the TESS and \textit{Kepler} results, we compare our ridge--savanna contrast with estimates derived from the discovery study and earlier \textit{Kepler} occurrence works, which use different planet samples and methodologies. We obtain the contrasts from the published occurrence rates or the weighted planet counts when the rates are unavailable, normalising each integrated occurrence by its period width and matching the period and radius selections to the available bins. Table~\ref{tab:historicalcontrasts} lists the intervals and corresponding contrasts. For the \textit{Kepler} estimates, we derive central 68\%
    uncertainty ranges by Monte Carlo propagation of the published errors or counting statistics, using a Poisson bootstrap for the weighted planet counts. We find that the central contrasts range from 1.6 to 2.2 and agree with the TESS measurement within their uncertainties. The approximately twofold excess over the adjacent savanna measured with TESS thus echoes the contrasts from the original ridge study and historical \textit{Kepler} occurrence works.

    The profile comparison provides a tentative indication of the ridge's morphology. In the main sample, a lognormal is preferred over a symmetric Gaussian, suggesting a narrower inner wing and a more extended outer wing, although the evidence remains weak ($\Delta\ln\mathcal Z_{\rm LN,G}=1.90$). The broken-power-law fit also hints at a steeper inner rise, although the difference between the slope magnitudes remains poorly constrained. Notably, such a shape would be compatible with HEM as a possible origin of the ridge. In the high-eccentricity limit, the minimum survival period is set mostly by planetary density \citep{Roche1849,MatsakosKonigl2016,CastroGonzalez2026}, so tidal disruption could produce a sharp decline in occurrence towards shorter periods. However, the outer extent of successfully circularised orbits also depends on planetary tidal dissipation, system age, and initial orbital conditions \citep[e.g.][]{OwenLai2018}. Therefore, variation in these conditions across systems could allow a more gradual outer decline. More precise occurrence measurements are nevertheless needed to establish whether the ridge is asymmetric, as well as whether its contrast varies across stellar types.

    Recently, \citet{Cui2026Ridge} reported an analysis of the same TESS-SPOC sample. Their model comparison likewise favours flexible profiles over a PL. They also find a limited advantage for PL+LN over BPL, which they interpret as indicating that a distinct ridge component is not required to explain the data. However, a BPL can itself describe a concentration when its slope changes sign, without an additional peak component. Indeed, a BPL is the preferred model for the hot Jupiter pileup based on both predictive performance and Bayesian evidence. For the ridge, our BPL fit gives a 99.7\% posterior probability of a peak ($\beta_1>0$, $\beta_2<0$) and places the break within the ridge at $p_b=3.73^{+0.73}_{-0.61}$~$\mathrm{d}$. Their BPL likewise peaks within the ridge at $p_b=4.27^{+0.69}_{-0.61}$~$\mathrm{d}$, with slopes $\beta_1=1.61^{+0.43}_{-0.36}$ and $\beta_2=-0.69^{+0.34}_{-0.38}$ in occurrence density. These results are compatible with ours despite the different period and radius selections. Therefore, the limited advantage of PL+LN over BPL does not imply that a ridge is unnecessary to explain the TESS data, given that the BPL fit itself describes a peak within the ridge that corresponds to the directly measured occurrence excess (Sect.~\ref{sec:results}).

    \section{Conclusions} \label{sec:conclusions}

    TESS provides an independent detection of the Neptunian ridge. For planets with radii of $\approx$~$5$--$8\,\mathrm{R}_\oplus$, the ridge's approximately twofold excess in occurrence density over the denser adjacent region has 99.9\% posterior support (one-sided Gaussian equivalent of $3.1\sigma$) and agrees in amplitude with estimates from the discovery study and earlier \textit{Kepler} occurrence works. Both predictive performance and Bayesian evidence strongly favour profiles peaking within the ridge over a single power law, with the asymmetric lognormal providing the preferred description, although the evidence for asymmetry remains weak. During its first two years, PLATO is expected to deliver a planet yield comparable to that of the TESS primary mission \citep{Cabrera2026}. The resulting occurrence measurements, combined with those from TESS and \textit{Kepler}, could help refine the ridge's morphology and explore possible differences across stellar types, providing new clues to how these planets reached their present orbits.

    \begin{acknowledgements} We thank the anonymous referee for constructive comments that improved the clarity of this Letter. We thank Michelle Kunimoto for helpful discussions on the Neptunian ridge in \textit{Kepler} and TESS occurrence data. This work has been carried out within the framework of the NCCR PlanetS supported by the Swiss National Science Foundation under grant 51NF40\_205606. \end{acknowledgements}

    \bibliographystyle{bibtex/aa} \bibliography{references}

\begin{thebibliography}{38}
\expandafter\ifx\csname natexlab\endcsname\relax\def\natexlab#1{#1}\fi

\bibitem[{Abhilash {et~al.}(2026)Abhilash, Bryan, \& Zanazzi}]{Abhilash2026}
Abhilash, S.~E., Bryan, M.~L., \& Zanazzi, J.~J. 2026, arXiv e-prints
  [\eprint[arXiv]{2611.35111}]

\bibitem[{Almenara {et~al.}(2026)Almenara, Castro-Gonz{\'a}lez, Mardling,
  Guillot, Bonfils, Thomas, Gauvrit, Abe, Agabi, Aubert, Bouchy, Cadieux,
  Ciardi, Deline, Deveny, D{\'i}az, Ehrenreich, Forveille, Heitzmann, Hirling,
  Hoogenboom, Howell, Leleu, Lendl, Littlefield, M{\'e}karnia, Pelletier,
  Psaridi, Rea, Song, Su{\'a}rez, Triaud, \& Udry}]{Almenara2026}
Almenara, J.~M., Castro-Gonz{\'a}lez, A., Mardling, R., {et~al.} 2026, \aap,
  714, A190

\bibitem[{Azzalini(1985)}]{Azzalini1985}
Azzalini, A. 1985, Scandinavian Journal of Statistics. Theory and Applications,
  12, 171

\bibitem[{{Bourrier} {et~al.}(2023){Bourrier}, {Attia}, {Mallonn}, {Marret},
  {Lendl}, {Konig}, {Krenn}, {Cretignier}, {Allart}, {Henry}, {Bryant},
  {Leleu}, {Nielsen}, {Hebrard}, {Hara}, {Ehrenreich}, {Seidel}, {dos Santos},
  {Lovis}, {Bayliss}, {Cegla}, {Dumusque}, {Boisse}, {Boucher}, {Bouchy},
  {Pepe}, {Lavie}, {Rey Cerda}, {S{\'e}gransan}, {Udry}, \&
  {Vrignaud}}]{Bourrier2023}
{Bourrier}, V., {Attia}, M., {Mallonn}, M., {et~al.} 2023, \aap, 669, A63

\bibitem[{{Bourrier} {et~al.}(2018){Bourrier}, {Lovis}, {Beust}, {Ehrenreich},
  {Henry}, {Astudillo-Defru}, {Allart}, {Bonfils}, {S{\'e}gransan}, {Delfosse},
  {Cegla}, {Wyttenbach}, {Heng}, {Lavie}, \& {Pepe}}]{Bourrier2018}
{Bourrier}, V., {Lovis}, C., {Beust}, H., {et~al.} 2018, \nat, 553, 477

\bibitem[{B{\"u}rkner {et~al.}(2021)B{\"u}rkner, Gabry, \&
  Vehtari}]{Buerkner2021}
B{\"u}rkner, P.-C., Gabry, J., \& Vehtari, A. 2021, Computational Statistics,
  36, 1243

\bibitem[{{Cabrera} {et~al.}(2026){Cabrera}, {Rauer}, {Samadi}, {Nascimbeni},
  {Boerner}, {Grie{\ss}bach}, {Paproth}, {Pertena{\i}s}, {Niemi}, {Csizmadia},
  {Abreu}, {Aerts}, {Aigrain}, {Ammler-von Eiff}, {Aparicio del Moral},
  {Appourchaux}, {Armstrong}, {Baeke}, {Balazs}, {Belkacem}, {Birch}, {Bluhm},
  {Boenke}, {Boquet}, {Bowling}, {Brown}, {Catala}, {Chaplin}, {Cunha},
  {Daminani}, {Davies}, {Davoult}, {De Angeli}, {De Ridder}, {Deleuil},
  {Desert}, {Diaz Garcia}, {Di Giorgio}, {Doyle}, {Edwards}, {Eigmueller},
  {Eising}, {Erikson}, {Nike Eschen}, {Ferrari}, {Ford},
  {Garc{\'\i}a-V{\'a}zquez}, {Gizon}, {Gomez Lopez}, {Gorius}, {Goupil},
  {Granata}, {Grenfell}, {Grolleau}, {Grziwa}, {Guillot}, {Harrison}, {Heller},
  {Heras}, {Hodgkin}, {Huygen}, {Jannsen}, {Kappel}, {Klagyivik}, {Koncz},
  {Kossakowska}, {Labiano}, {Lam}, {Lanza}, {Lendl}, {Levillain}, {Lobon
  Villanueva}, {Magrin}, {Malavolta}, {Marinoni}, {Marrese}, {Garcia}, {Mas
  Hesse}, {Maxted}, {McCormac}, {Miglio}, {Montalto}, {Morel}, {Morena},
  {Moya}, {Munari}, {Nielsen}, {Ouazzani}, {Pagano}, {Pastor Morales}, {Peter},
  {Philidet}, {Piotto}, {Plasson}, {Pollacco}, {Puga}, {Ragazzoni}, {Ramos
  Zapata}, {Regibo}, {Rixon}, {Robles Mu{\~n}oz}, {Rodriguez Gomez}, {Royer},
  {Sanchez Carrasco}, {Sanz Mesa}, {Schwarzkopf}, {Seynaeve}, {Smith}, {Smith},
  {Smith}, {Sulis}, {Talens}, {Titz-Weider}, {Udry}, {Vandenbussche},
  {Valtchanov}, {Verhoeve}, {Walton}, {Walton}, {Wilson}, {Witteck}, {Wolter},
  {Ziemke}, \& {Zwintz}}]{Cabrera2026}
{Cabrera}, J., {Rauer}, H., {Samadi}, R., {et~al.} 2026, arXiv e-prints,
  arXiv:2604.04818

\bibitem[{{Castro-Gonz{\'a}lez}
  {et~al.}(2026{\natexlab{a}}){Castro-Gonz{\'a}lez}, {Barrag{\'a}n},
  {Armstrong}, {Aguichine}, {Bourrier}, {Ehrenreich}, {Tao}, {Lillo-Box},
  {Standing}, {Sousa}, {Delgado-Mena}, {Moya}, {Adibekyan}, {Lendl}, {Hellier},
  {Howell}, {Furlan}, {Ziegler}, {Correia}, {Cui}, {Figueira}, {Jenkins},
  {Fetzner Keniger}, {Mer{\'\i}n}, {Osborn}, {Parc}, {Pepe}, {Wheatley}, \&
  {Winn}}]{CastroGonzalez2026TOI426}
{Castro-Gonz{\'a}lez}, A., {Barrag{\'a}n}, O., {Armstrong}, D.~J., {et~al.}
  2026{\natexlab{a}}, \aap, 714, A425

\bibitem[{{Castro-Gonz{\'a}lez}
  {et~al.}(2026{\natexlab{b}}){Castro-Gonz{\'a}lez}, {Bourrier}, {Ehrenreich},
  {Armstrong}, {Correia}, \& {Lendl}}]{CastroGonzalez2026}
{Castro-Gonz{\'a}lez}, A., {Bourrier}, V., {Ehrenreich}, D., {et~al.}
  2026{\natexlab{b}}, \aap, 709, L17

\bibitem[{{Castro-Gonz{\'a}lez}
  {et~al.}(2024{\natexlab{a}}){Castro-Gonz{\'a}lez}, {Bourrier}, {Lillo-Box},
  {Delisle}, {Armstrong}, {Barrado}, \& {Correia}}]{CastroGonzalez2024}
{Castro-Gonz{\'a}lez}, A., {Bourrier}, V., {Lillo-Box}, J., {et~al.}
  2024{\natexlab{a}}, \aap, 689, A250

\bibitem[{{Castro-Gonz{\'a}lez}
  {et~al.}(2024{\natexlab{b}}){Castro-Gonz{\'a}lez}, {Lillo-Box}, {Armstrong},
  {Acu{\~n}a}, {Aguichine}, {Bourrier}, {Gandhi}, {Sousa}, {Delgado-Mena},
  {Moya}, {Adibekyan}, {Correia}, {Barrado}, {Damasso}, {Winn}, {Santos},
  {Barkaoui}, {Barros}, {Benkhaldoun}, {Bouchy}, {Brice{\~n}o}, {Caldwell},
  {Collins}, {Essack}, {Ghachoui}, {Gillon}, {Hounsell}, {Jehin}, {Jenkins},
  {Keniger}, {Law}, {Mann}, {Nielsen}, {Pozuelos}, {Schanche}, {Seager}, {Tan},
  {Timmermans}, {Villase{\~n}or}, {Watkins}, \&
  {Ziegler}}]{CastroGonzalez2024TOI5005}
{Castro-Gonz{\'a}lez}, A., {Lillo-Box}, J., {Armstrong}, D.~J., {et~al.}
  2024{\natexlab{b}}, \aap, 691, A233

\bibitem[{{Correia} {et~al.}(2020){Correia}, {Bourrier}, \&
  {Delisle}}]{Correia2020}
{Correia}, A.~C.~M., {Bourrier}, V., \& {Delisle}, J.-B. 2020, \aap, 635, A37

\bibitem[{{Cui} {et~al.}(2026{\natexlab{a}}){Cui}, {Armstrong},
  {Hadjigeorghiou}, {Lafarga}, {Kunovac}, {Doyle}, {Nieto}, \&
  {D{\'\i}az}}]{Cui2026}
{Cui}, K., {Armstrong}, D.~J., {Hadjigeorghiou}, A., {et~al.}
  2026{\natexlab{a}}, \mnras, 546, stag022

\bibitem[{{Cui} {et~al.}(2026{\natexlab{b}}){Cui}, {Armstrong},
  {Hadjigeorghiou}, {Lafarga}, {Osborn}, {King}, {Hallatt}, {Kunovac}, \&
  {Doyle}}]{Cui2026Ridge}
{Cui}, K., {Armstrong}, D.~J., {Hadjigeorghiou}, A., {et~al.}
  2026{\natexlab{b}}, \mnras, stag2018

\bibitem[{{Ehrenreich} \& {D{\'e}sert}(2011)}]{EhrenreichDesert2011}
{Ehrenreich}, D. \& {D{\'e}sert}, J.~M. 2011, \aap, 529, A136

\bibitem[{{Fabrycky} \& {Tremaine}(2007)}]{FabryckyTremaine2007}
{Fabrycky}, D. \& {Tremaine}, S. 2007, \apj, 669, 1298

\bibitem[{{Fressin} {et~al.}(2013){Fressin}, {Torres}, {Charbonneau}, {Bryson},
  {Christiansen}, {Dressing}, {Jenkins}, {Walkowicz}, \&
  {Batalha}}]{Fressin2013}
{Fressin}, F., {Torres}, G., {Charbonneau}, D., {et~al.} 2013, \apj, 766, 81

\bibitem[{{Hallatt} \& {Millholland}(2026)}]{HallattMillholland2026}
{Hallatt}, T. \& {Millholland}, S. 2026, \apj, 997, 138

\bibitem[{{Howard} {et~al.}(2012){Howard}, {Marcy}, {Bryson}, {Jenkins},
  {Rowe}, {Batalha}, {Borucki}, {Koch}, {Dunham}, {Gautier}, {Van Cleve},
  {Cochran}, {Latham}, {Lissauer}, {Torres}, {Brown}, {Gilliland}, {Buchhave},
  {Caldwell}, {Christensen-Dalsgaard}, {Ciardi}, {Fressin}, {Haas}, {Howell},
  {Kjeldsen}, {Seager}, {Rogers}, {Sasselov}, {Steffen}, {Basri},
  {Charbonneau}, {Christiansen}, {Clarke}, {Dupree}, {Fabrycky}, {Fischer},
  {Ford}, {Fortney}, {Tarter}, {Girouard}, {Holman}, {Johnson}, {Klaus},
  {Machalek}, {Moorhead}, {Morehead}, {Ragozzine}, {Tenenbaum}, {Twicken},
  {Quinn}, {Isaacson}, {Shporer}, {Lucas}, {Walkowicz}, {Welsh}, {Boss},
  {Devore}, {Gould}, {Smith}, {Morris}, {Prsa}, {Morton}, {Still}, {Thompson},
  {Mullally}, {Endl}, \& {MacQueen}}]{Howard2012}
{Howard}, A.~W., {Marcy}, G.~W., {Bryson}, S.~T., {et~al.} 2012, \apjs, 201, 15

\bibitem[{Huang {et~al.}(2016)Huang, Wu, \& Triaud}]{Huang2016}
Huang, C., Wu, Y., \& Triaud, A. H. M.~J. 2016, \apj, 825, 98

\bibitem[{{Kunimoto} \& {Matthews}(2020)}]{Kunimoto2020}
{Kunimoto}, M. \& {Matthews}, J.~M. 2020, \aj, 159, 248

\bibitem[{{Matsakos} \& {K{\"o}nigl}(2016)}]{MatsakosKonigl2016}
{Matsakos}, T. \& {K{\"o}nigl}, A. 2016, \apjl, 820, L8

\bibitem[{{Mulders} {et~al.}(2015){Mulders}, {Pascucci}, \&
  {Apai}}]{Mulders2015}
{Mulders}, G.~D., {Pascucci}, I., \& {Apai}, D. 2015, \apj, 814, 130

\bibitem[{Mustill {et~al.}(2015)Mustill, Davies, \& Johansen}]{Mustill2015}
Mustill, A.~J., Davies, M.~B., \& Johansen, A. 2015, \apj, 808, 14

\bibitem[{Nelsen(2006)}]{Nelsen2006}
Nelsen, R.~B. 2006, {An Introduction to Copulas}, 2nd edn., Springer Series in
  Statistics (New York, NY: Springer)

\bibitem[{{Owen} \& {Lai}(2018)}]{OwenLai2018}
{Owen}, J.~E. \& {Lai}, D. 2018, \mnras, 479, 5012

\bibitem[{{Ricker} {et~al.}(2015){Ricker}, {Winn}, {Vanderspek}, {Latham},
  {Bakos}, {Bean}, {Berta-Thompson}, {Brown}, {Buchhave}, {Butler}, {Butler},
  {Chaplin}, {Charbonneau}, {Christensen-Dalsgaard}, {Clampin}, {Deming},
  {Doty}, {De Lee}, {Dressing}, {Dunham}, {Endl}, {Fressin}, {Ge}, {Henning},
  {Holman}, {Howard}, {Ida}, {Jenkins}, {Jernigan}, {Johnson}, {Kaltenegger},
  {Kawai}, {Kjeldsen}, {Laughlin}, {Levine}, {Lin}, {Lissauer}, {MacQueen},
  {Marcy}, {McCullough}, {Morton}, {Narita}, {Paegert}, {Palle}, {Pepe},
  {Pepper}, {Quirrenbach}, {Rinehart}, {Sasselov}, {Sato}, {Seager},
  {Sozzetti}, {Stassun}, {Sullivan}, {Szentgyorgyi}, {Torres}, {Udry}, \&
  {Villasenor}}]{Ricker2015}
{Ricker}, G.~R., {Winn}, J.~N., {Vanderspek}, R., {et~al.} 2015, Journal of
  Astronomical Telescopes, Instruments, and Systems, 1, 014003

\bibitem[{Roche(1849)}]{Roche1849}
Roche, {\'E}. 1849, M{\'e}moires de l'Acad{\'e}mie des Sciences et Lettres de
  Montpellier, Section des Sciences, 1, 243

\bibitem[{{Speagle}(2020)}]{Speagle2020}
{Speagle}, J.~S. 2020, \mnras, 493, 3132

\bibitem[{{Szab{\'o}} \& {Kiss}(2011)}]{SzaboKiss2011}
{Szab{\'o}}, G.~M. \& {Kiss}, L.~L. 2011, \apjl, 727, L44

\bibitem[{{Thomas} {et~al.}(2026){Thomas}, {Nielsen}, {Cridland}, {Schlecker},
  \& {Vach}}]{Thomas2026}
{Thomas}, L., {Nielsen}, L.~D., {Cridland}, A.~J., {Schlecker}, M., \& {Vach},
  S. 2026, \aap, in press

\bibitem[{{Udry} {et~al.}(2003){Udry}, {Mayor}, \& {Santos}}]{Udry2003}
{Udry}, S., {Mayor}, M., \& {Santos}, N.~C. 2003, \aap, 407, 369

\bibitem[{Vehtari {et~al.}(2017)Vehtari, Gelman, \& Gabry}]{Vehtari2017}
Vehtari, A., Gelman, A., \& Gabry, J. 2017, Statistics and Computing, 27, 1413

\bibitem[{Wu {et~al.}(2023)Wu, Rice, \& Wang}]{WuRiceWang2023}
Wu, D.-H., Rice, M., \& Wang, S. 2023, \aj, 165, 171

\bibitem[{{Wu} {et~al.}(2007){Wu}, {Murray}, \& {Ramsahai}}]{Wu2007}
{Wu}, Y., {Murray}, N.~W., \& {Ramsahai}, J.~M. 2007, \apj, 670, 820

\bibitem[{{Zanazzi} {et~al.}(2026){Zanazzi}, {MacLeod}, {Bryan}, \&
  {Mahadevan}}]{Zanazzi2026}
{Zanazzi}, J.~J., {MacLeod}, M., {Bryan}, M.~L., \& {Mahadevan}, S. 2026, \apj,
  1010, 185

\bibitem[{Zink {et~al.}(2023)Zink, Hardegree-Ullman, Christiansen, Petigura,
  Boley, Bhure, Rice, Yee, Isaacson, Fernandes, Howard, Blunt, Lubin, Chontos,
  Pidhorodetska, \& MacDougall}]{Zink2023}
Zink, J.~K., Hardegree-Ullman, K.~K., Christiansen, J.~L., {et~al.} 2023, \aj,
  165, 262

\bibitem[{Zink \& Howard(2023)}]{ZinkHoward2023}
Zink, J.~K. \& Howard, A.~W. 2023, \apjl, 956, L29

\end{thebibliography}

    \appendix \nolinenumbers  \section{Models and statistical comparison} \label{app:models}

    \subsection{Occurrence models and priors}

    We consider six models to describe the occurrence density, $h(P)$, across $D$, $R$, and $S$. The PL has the shape \begin{equation} u(P)=P^\beta, \end{equation} while the BPL is defined by \begin{equation} u(P)= \begin{cases} (P/p_b)^{\beta_1}, & P<p_b,\\ (P/p_b)^{\beta_2}, & P\geq p_b \end{cases}, \end{equation} where $\beta$ is the PL exponent, $\beta_1$ and $\beta_2$ are the inner and outer BPL exponents, and $p_b$ is the break period. For the Gaussian, \begin{equation} v(P)=\exp\left[-\frac{(P-\mu)^2}{2\sigma^2}\right], \end{equation} and for the lognormal, \begin{equation} v(P)=P^{-1} \exp\left[-\frac{[\ln(P/\mu)-\tau^2]^2}{2\tau^2}\right], \end{equation} where $\mu$ denotes the Gaussian centre or the lognormal mode, while $\sigma$ and $\tau$ set their widths in period and log period, respectively. For these four models, we divide each shape by its integral over $D$, $R$, and $S$, and then multiply by the total occurrence $T$ to obtain $h(P)$. For PL+G and PL+LN, we combine the PL background $u(P)$ with the corresponding peak $v(P)$: \begin{equation} h(P)=T\left[(1-f)\frac{u(P)}{\int u(P)\,\mathrm dP} +f\frac{v(P)}{\int v(P)\,\mathrm dP}\right], \end{equation} where both integrals cover $D$, $R$, and $S$, and $f$ gives the fraction of total occurrence in the peak. To compare with the published rates, we predict the occurrence in period bin $j$ for model parameters $\theta$ as $F_j(\theta)=\int_{P_j}^{P_{j+1}}h(P;\theta)\,\mathrm dP$.

    For the BPL, we sample the mean slope $\beta_m=(\beta_1+\beta_2)/2$ and the slope change $\delta=\beta_1-\beta_2$, with $\delta=0$ recovering the PL. We adopt uniform priors $\beta,\beta_m\sim\mathcal U(-6,6)$, $\delta\sim\mathcal U(-12,12)$, $f\sim\mathcal U(0,1)$, and log-uniform priors $T\sim\mathcal{LU}(10^{-4},10)$ and $\sigma/\mu,\tau\sim\mathcal{LU}(0.04,2)$. For the break and peak location, we use $p_b\sim\mathcal{LU}(P_{\min},P_{\max})$ and $\mu\sim\mathcal U(P_{\min},P_{\max})$, where $P_{\min}$ and $P_{\max}$ delimit the period range spanning $D$, $R$, and $S$. These priors allow either slope sign and a peak anywhere in the period range. We assess sensitivity to these priors in Appendix~\ref{app:priors}.

    \subsection{Likelihood function}

    We examine the occurrence posteriors and find departures from a Gaussian shape. We therefore compare several joint distributions to describe their uncertainties and correlations between bins. We fit each distribution to 80\% of the posterior samples and evaluate it on the remaining 20\%. Across all stellar types and radius ranges, a Gaussian copula with skew-normal marginals \citep{Azzalini1985,Nelsen2006} improves on a multivariate Gaussian. We also test more elaborate marginals (i.e., Gaussian mixtures, sinh--arcsinh, and skew-$t$) and find at most modest improvements, so we adopt the simpler skew-normal representation. We fit each marginal to all posterior samples by maximum likelihood and estimate the copula correlations from the Gaussian-transformed samples. We denote the fitted joint density of the logarithms of the integrated bin occurrences by $q$ and define the likelihood as \begin{equation} \mathcal L_{\rm rates}(\theta) =q[\ln\boldsymbol F(\theta)], \end{equation} where $\boldsymbol F(\theta)$ collects the predicted occurrences in the fitted bins.

    \subsection{Leave-one-out comparison} \label{app:loo}

    We use LOO to assess how well each model predicts an omitted bin. We omit one bin $i$ at a time and fit each model to the remaining bins using their joint uncertainty distribution, $q_{-i}$. We obtain the parameter posterior $p_{-i}(\theta\mid M)$ through nested sampling with \texttt{dynesty} \citep{Speagle2020}, using 1000 live points and a stopping tolerance of 0.025 in log evidence. We then evaluate the omitted bin conditional on those retained. Writing $\boldsymbol x(\theta)=\ln\boldsymbol F(\theta)$, the conditional density is $q_{i\mid-i}(\boldsymbol x)=q(\boldsymbol x)/q_{-i}(\boldsymbol x_{-i})$ \citep[Eq.~3]{Buerkner2021}. Following \citet[Eq.~1]{Buerkner2021}, we average this density over $p_{-i}(\theta\mid M)$ and take the logarithm to obtain \begin{equation} \ell_{i,M}=\ln\int q_{i\mid-i}[\boldsymbol x(\theta)] p_{-i}(\theta\mid M)\,\mathrm d\theta. \end{equation} For each model $M$, we estimate the expected log predictive density as $\mathrm{ELPD}_M=\sum_i\ell_{i,M}$ \citep{Vehtari2017}. We report differences between models with the descriptive error $[\sum_i(d_i-\bar d)^2]^{1/2}$, where $d_i=\ell_{i,A}-\ell_{i,B}$ is the difference for bin $i$. We also fit each model to all bins to estimate its parameters and uncertainties, and plot the highest-likelihood sampled profile.

    \subsection{Bayesian evidence comparison}

    We compare the six models through their Bayesian evidence, the likelihood averaged over the parameter prior: \begin{equation} \mathcal Z_M=\int\mathcal L_{\rm rates}(\theta) \pi_M(\theta)\,\mathrm d\theta. \end{equation} We evaluate this integral with \texttt{dynesty} by using the same number of live points and stopping tolerance as for the LOO comparison. We run each model with five independent seeds and report the mean log-evidence differences. To estimate the numerical uncertainty of each mean, we take the larger of its propagated integration error and its standard error across seeds. We obtain the model parameters by averaging the posteriors from the five runs.

    \subsection{Sensitivity to priors} \label{app:priors}

    We repeat the analysis with two narrower and two wider choices of priors to assess how these ranges affect the model comparison. The narrower sets use $T\sim\mathcal{LU}(0.001,2)$ and $\mathcal{LU}(0.003,1)$, $\beta,\beta_m\sim\mathcal U(-4,4)$ and $\mathcal U(-3,3)$, $\delta\sim\mathcal U(-8,8)$ and $\mathcal U(-6,6)$, and $\sigma/\mu,\tau\sim\mathcal{LU}(0.08,1.2)$ and $\mathcal{LU}(0.12,0.8)$, respectively. For the two wider sets, we broaden the slope priors to $\beta,\beta_m\sim\mathcal U(-9,9)$ and $\mathcal U(-12,12)$, and $\delta\sim\mathcal U(-18,18)$ and $\mathcal U(-24,24)$, with $\sigma/\mu,\tau\sim\mathcal{LU}(0.02,3)$ and $\mathcal{LU}(0.01,4)$, respectively. In all four tests, we retain the broad priors on the break period, peak location, and mixture fraction.

    BPL remains strongly favoured over PL, with $\Delta\ln\mathcal Z_{\rm BPL,PL}=5.76$--$7.13$. LN retains the highest evidence and PL the lowest, although the strength of the preference for LN over BPL depends on the prior ranges. Its log-evidence advantage varies from 0.97 to 2.67, reaching moderate support only with the widest priors. The comparison with the symmetric G varies little, with $\Delta\ln\mathcal Z_{\rm LN,G}=1.90$--$2.10$, giving weak evidence for asymmetry. The LOO comparison reinforces the departure from a single PL, with BPL improving on PL by 10.97--12.91 in ELPD. LN retains the highest predictive score throughout and improves on G by 4.34--4.71. The fitted peak locations also remain stable, with median BPL breaks at 3.71--3.73~$\mathrm{d}$ and median LN modes at 4.34--4.37~$\mathrm{d}$, both within the ridge. The peak therefore persists across the tested priors, while its asymmetry remains tentative.

    \onecolumn \raggedbottom \begin{multicols}{2}   \section{The ridge across spectral types} \label{app:spectral}

    We compare the ridge occurrence density with that of the denser neighbouring region separately for F, G, and K stars. Over 5.02--7.96~$\mathrm{R}_\oplus$, this gives $J=1.20^{+0.56}_{-0.38}$ for F stars, $J=1.25^{+0.43}_{-0.33}$ for G stars, and $J=2.78^{+1.27}_{-0.85}$ for K stars, with corresponding posterior probabilities $p_+=68.319\%$, 76.379\%, and 99.727\%, respectively. The median contrasts hint at a ridge excess for each stellar type. However, no individual sample reaches $3\sigma$ for $J>1$. Posterior support remains weak for F and G stars, although it approaches $3\sigma$ for K stars, which show a higher median contrast. Notably, despite their different median values, the central 68\% credible intervals of all three stellar samples overlap with that of FGK. Similarly, although pairwise comparisons suggest a larger contrast for K stars than for F or G stars, support for this difference reaches at most $1.68\sigma$. Overall, we find no significant difference in ridge contrast between stellar types. Figure~\ref{fig:spectralprofiles} shows the occurrence profiles by stellar type and radius range.

    Limited precision may explain why the ridge is not detected at $3\sigma$ in the separate stellar samples. We test this possibility with injection--recovery experiments. We use a fixed reference profile based on the FGK occurrence rates with $J_0=2.14$. We simulate measurements with errors drawn from the joint uncertainty distributions in the posteriors, thus preserving the precision of each sample and the correlations between regions. The ridge excess reaches $Z_{\rm eq}\geq3$ in 40.6\% of FGK experiments, compared with 6.0\%, 15.3\%, and 5.5\% for F, G, and K stars, respectively. Recovery rates remain low for injected contrasts comparable to the FGK reference, indicating that individual $3\sigma$ detections are unlikely at the precision of the separate stellar samples.

    \end{multicols}

    \begin{appendixfigure} \includegraphics[width=0.97\textwidth]{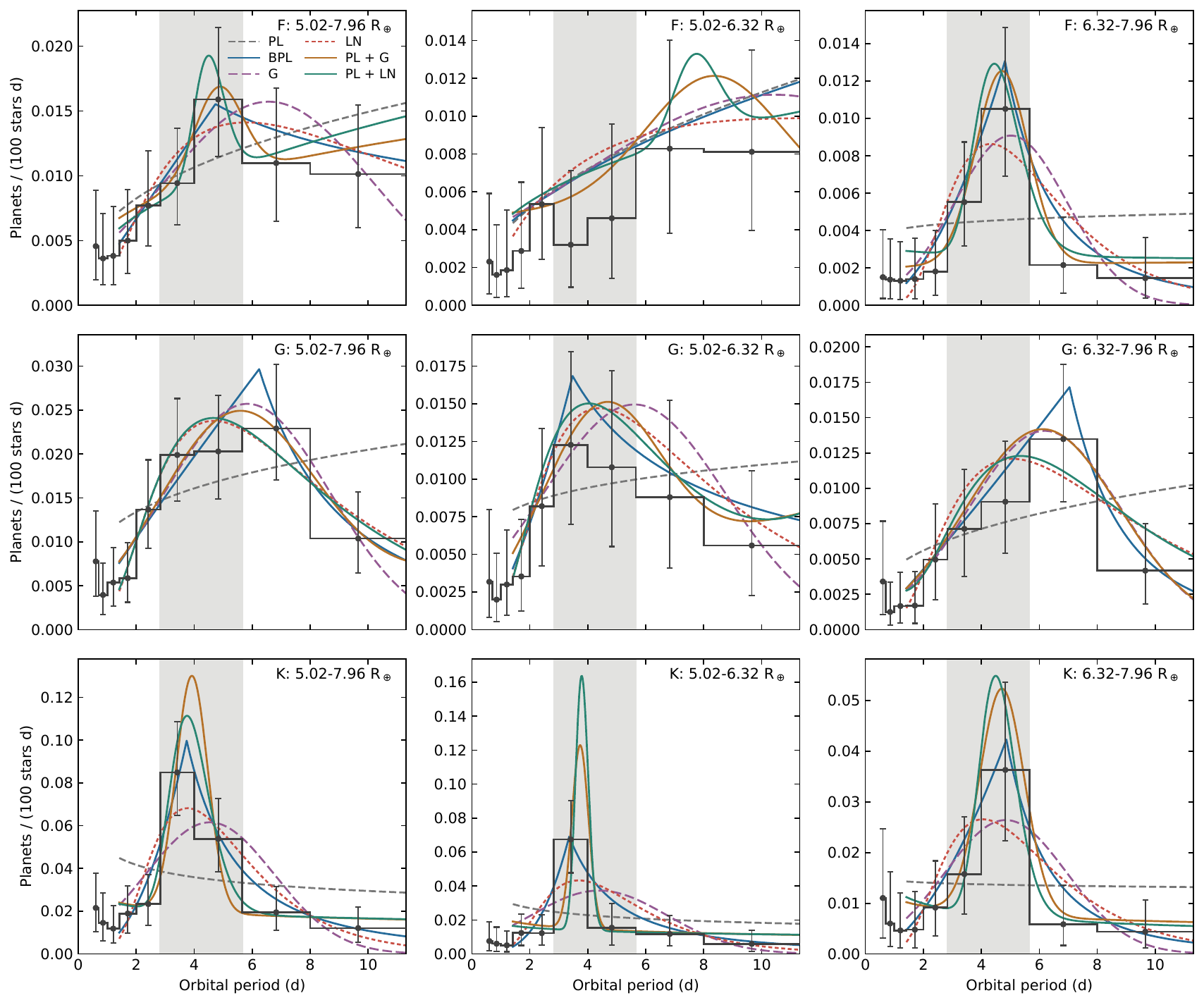} \caption{Same as Fig.~\ref{fig:fgk}, but for F, G, and K stars (top, middle, and bottom rows, respectively).} \label{fig:spectralprofiles} \end{appendixfigure}

    \begin{multicols}{2}   \section{Other radius intervals} \label{app:populations}

    We extend the analysis to gas giants, sub-Neptunes, and the two transition regions introduced in Sect.~\ref{sec:samples}. We show their occurrence profiles in Fig.~\ref{fig:populations}. To investigate whether concentrations at periods similar to those of the ridge are present, we consider the same period windows and compare the occurrence density in $R$ with those in $D$ and $S$.

    For gas giants (10.02--15.89~$\mathrm{R}_\oplus$), we find a pronounced central concentration corresponding to the well-known hot Jupiter pileup. Its occurrence density is approximately twice that of the denser neighbouring region ($J=2.31^{+0.30}_{-0.26}$), and the exterior contrast $C$ is nearly identical. The central region has a higher occurrence density than both neighbouring regions in all 6960 posterior samples ($J>1$). To describe the pileup's shape, we compare the six occurrence models using the procedure in Appendix~\ref{app:models}. We find that BPL has the highest evidence and predictive score, with $\Delta\ln\mathcal Z_{\rm BPL,PL}=57.79$ and $\Delta\mathrm{ELPD}_{\rm BPL,PL}=90.33\pm37.10$. Its advantage over the closest alternative, PL+LN, is moderate, with $\Delta\ln\mathcal Z_{\rm BPL,PL+LN}=2.60$. The BPL places the maximum at $p_b=3.24^{+0.19}_{-0.17}$~$\mathrm{d}$, with a rising inner slope $\beta_1=3.68^{+0.68}_{-0.59}$ and a falling outer slope $\beta_2=-1.32^{+0.19}_{-0.20}$. In PL+LN, the lognormal component peaks at $\mu=3.68^{+0.13}_{-0.11}$~$\mathrm{d}$, also within $R$. Compared with the ridge, the BPL gives a similar break period and median outer slope, with a steeper median inner rise. However, the central 68\% credible intervals overlap for all three parameters.

    For sub-Neptune planets (2.00--3.99~$\mathrm{R}_\oplus$), we obtain $J=C=0.58^{+0.10}_{-0.09}$, where the two contrasts coincide because the $S$ region is the denser neighbour in every sample. These values favour a lower occurrence density in the central interval $R$ than in $S$, opposite to the excess seen in the Neptunian ridge and hot Jupiter pileup. Only two of the 6960 posterior samples have $J>1$, giving 99.971\% posterior support for $J<1$, equivalent to $3.4\sigma$. The sub-Neptune measurements are therefore precise enough to rule out an occurrence excess in the $R$ period window. We also find comparable Bayesian evidence for the six models, with all log-evidence differences below 2.5 and no clearly preferred model.

    We also examine the 3.99--5.02 and 7.96--10.02~$\mathrm{R}_\oplus$ transition regions between the adopted ridge selection and the sub-Neptune and gas giant populations. These correspond to the frontier Neptunian regimes examined by \citet{CastroGonzalez2024}, with boundaries adapted to the available cells. For the lower and upper radius intervals, $J=0.58^{+0.29}_{-0.21}$ and $J=0.59^{+0.21}_{-0.17}$, with posterior probabilities of $J>1$ of 8.578\% and 3.563\%, respectively. The exterior contrasts $C$ are nearly identical to $J$, as $S$ is again the denser neighbour in almost every posterior sample. The median contrasts do not suggest an occurrence excess in $R$, but posterior support for $J<1$ reaches only $1.37\sigma$ and $1.80\sigma$, respectively. The larger uncertainties therefore prevent us from excluding an occurrence excess in $R$ at $3\sigma$, leaving its presence in these transition regions unresolved. We likewise find no clear Bayesian preference among the six models in either interval. \end{multicols}

    \begin{appendixfigure} \includegraphics[width=0.94\textwidth]{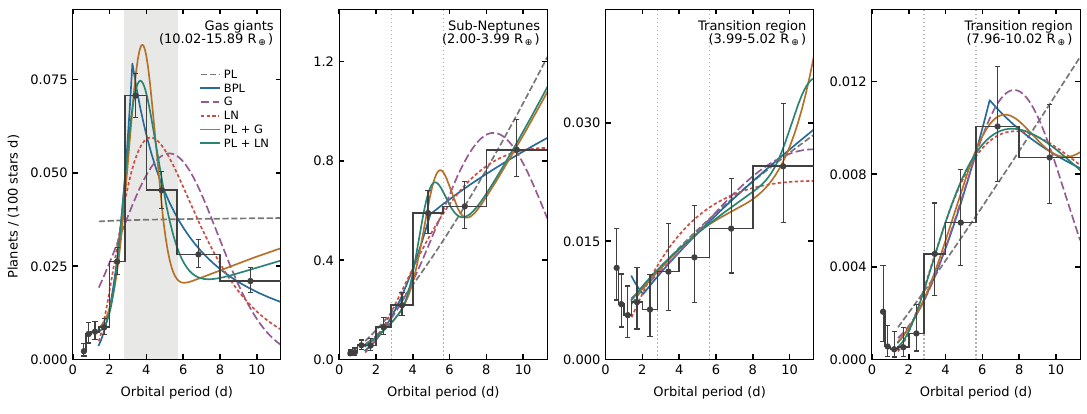} \caption{Same as Fig.~\ref{fig:fgk}, but for gas giants, sub-Neptunes, and the two transition regions. The shaded band for gas giants and dotted lines in the other panels delimit the adopted 2.83--5.66~$\mathrm{d}$ central comparison window $R$.} \label{fig:populations} \end{appendixfigure}

    \section{Additional tables} \label{app:comparisontables}

    \begin{multicols}{2} \raggedcolumns

    \begin{appendixtable} \caption{LOO and Bayesian evidence comparisons across the $D$, $R$, and $S$ regions for 5.02--7.96~$\mathrm{R}_\oplus$ planets.\strut} \label{tab:profilecomparison} \centering\small \setlength{\tabcolsep}{6pt} \renewcommand{\arraystretch}{1.4} \begin{tabular*}{\linewidth}{@{\extracolsep{\fill}}lcc@{}} \hline\hline Model & $\Delta\mathrm{ELPD}$ (LOO) & $\Delta\ln\mathcal Z_{M,{\rm BPL}}$ \\ \midrule Power law (PL) & $-12.07\pm4.28$ & $-6.41\pm0.06$ \\ Broken power law (BPL) & $0$ & $0$ \\ Gaussian (G) & $-3.36\pm1.67$ & $-0.16\pm0.07$ \\ Lognormal (LN) & $1.14\pm0.41$ & $1.74\pm0.08$ \\ PL+G & $-1.58\pm0.95$ & $-0.36\pm0.06$ \\ PL+LN & $0.24\pm0.59$ & $0.29\pm0.06$ \\ \bottomrule \end{tabular*} \end{appendixtable}

    \columnbreak

    \begin{appendixtable} \caption{Ridge--savanna occurrence contrasts based on the data from different studies.\strut} \label{tab:historicalcontrasts} \centering\small \setlength{\tabcolsep}{2pt} \renewcommand{\arraystretch}{1.4} \begin{tabular*}{\linewidth}{@{\extracolsep{\fill}}ccccc@{}} \hline\hline Radius ($\mathrm{R}_\oplus$) & $R$ ($\mathrm{d}$) & $S$ ($\mathrm{d}$) & $C$ & Reference \\ \midrule 5.66--8.00 & 3.42--5.85 & 5.85--10.00 & $1.58^{+1.11}_{-0.74}$ & 1 \\ 4.00--6.00 & 3.42--5.85 & 5.85--10.00 & $2.07^{+1.20}_{-0.71}$ & 2 \\ 5.50--8.50 & 3.20--5.70 & 5.70--13.20 & $1.65^{+0.64}_{-0.47}$ & 3 \\ 4.00--8.00 & 3.13--6.25 & 6.25--12.50 & $2.08^{+1.62}_{-0.88}$ & 4 \\ 5.50--8.50 & 3.20--5.70 & 5.70--13.20 & $2.16^{+0.99}_{-0.73}$ & 5 \\ 5.02--7.96 & 2.83--5.66 & 5.66--11.31 & $2.14^{+0.64}_{-0.49}$ & 6 \\ \bottomrule \end{tabular*} \tablebib{(1) \citet{Howard2012}; (2) \citet{Fressin2013}; (3) \citet{Mulders2015}; (4) \hyperlink{cite.Kunimoto2020}{Kunimoto \& Matthews} \citeyearpar{Kunimoto2020}; (5) \hyperlink{cite.CastroGonzalez2024}{Castro-Gonz\'alez et al.} \citeyearpar{CastroGonzalez2024}; (6) \citet{Cui2026}.} \end{appendixtable}

    \end{multicols}

    \clearpage

    \section{Ridge and pileup occurrence per unit logarithmic period} \label{app:logperiod}

    \begin{appendixfigure} \includegraphics[width=\textwidth]{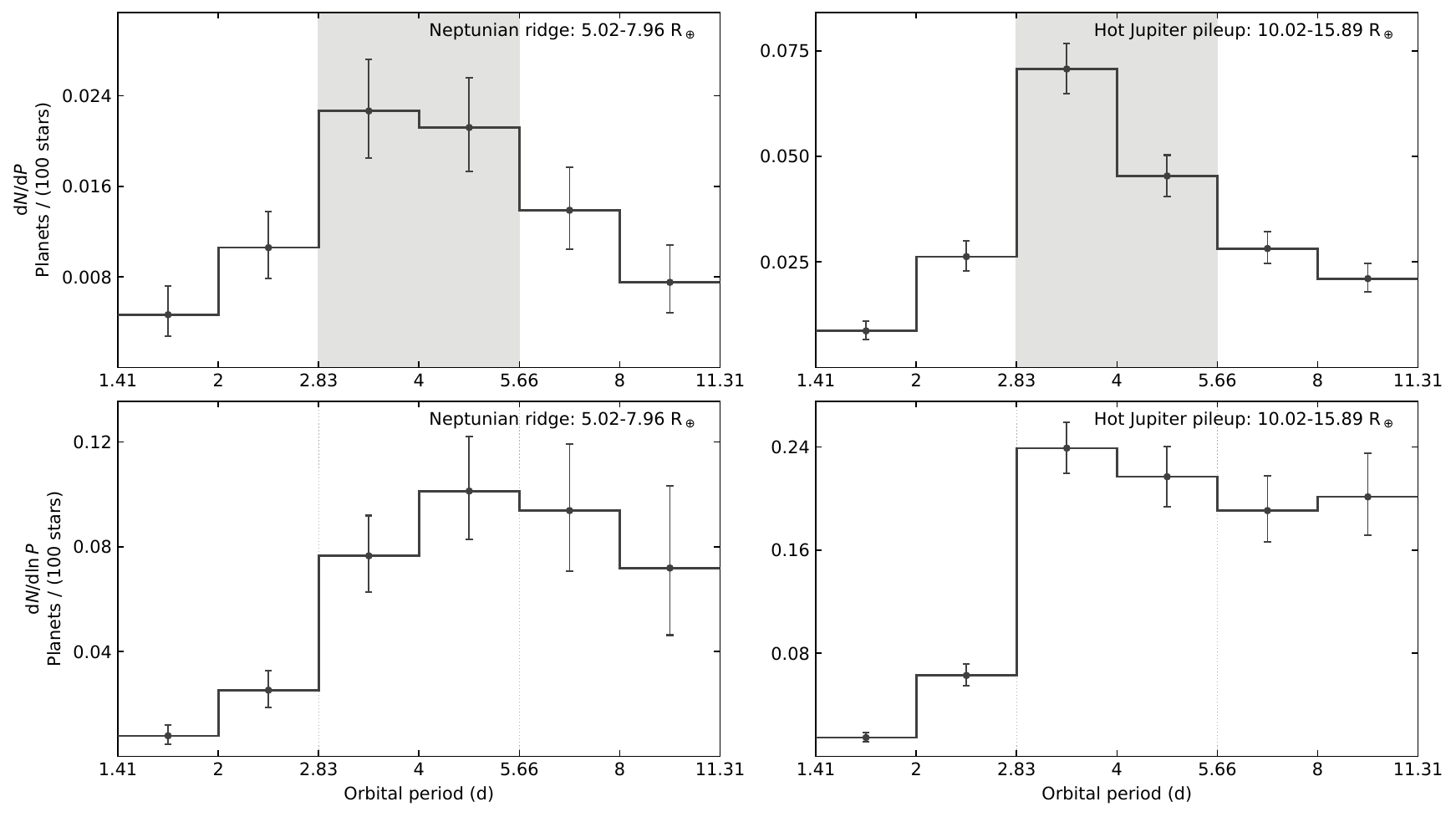} \caption{Same as Fig.~\ref{fig:fgk}, but for the ridge (left) and the pileup (right), with occurrence density expressed per day (top) and per unit $\ln P$ (bottom). Dotted lines in the lower panels mark the same per-day excess interval shaded above.} \label{fig:logperiod} \end{appendixfigure}

    \begin{multicols}{2}

    Figure~\ref{fig:logperiod} shows the flattening of the ridge and pileup profiles when occurrence is expressed per unit $\ln P$, with no statistically significant excess over the exterior region (one-sided Gaussian posterior equivalents of $0.25\sigma$ for the ridge and $1.19\sigma$ for the pileup), consistent with the profiles reported in previous occurrence studies \citep[e.g.][]{Kunimoto2020,Zink2023,ZinkHoward2023}. We emphasise that expressing the same occurrence data through a different density does not invalidate the per-day excesses by which these features were originally defined, whose relevance is reinforced by systematic differences in physical properties and orbital architectures between planets within and just beyond each overdensity (Sect.~\ref{sec:introduction}).

    \end{multicols}

    \end{document}